\documentclass[aip,amsmath,preprint,reprint,longbibliography]{revtex4-1}
\usepackage{mmap}
\usepackage{hyperref}
\usepackage[usenames]{xcolor}

\DeclareMathOperator{\erf}{erf}
\DeclareMathOperator{\erfc}{erfc}
\DeclareMathOperator{\erfcx}{erfcx}
\newcommand{\de}{\,\mathrm{d}}
\newcommand{\ts}[1]{{\scriptsize(#1)}}

\definecolor{mark}{rgb}{0.75,0.0,0.0}
\newcommand{\mm}[1]{#1}

\begin{document}

\title[Global approximations to the error function]
{Global approximations to the error function of real argument\\ for vectorized computation}

\author{Dimitri N. Laikov}
\email[]{dimitri\_laikov@mail.ru}
\affiliation{Chemistry Department, Moscow State University, 119991 Moscow, Russia}


\date{\today}

\begin{abstract}
The error function of real argument can be approximated to a given uniform relative accuracy
by a single closed-form expression for the whole variable range
either in terms of addition, multiplication, division, and square root operations only,
or also using the exponential function.
\mm{Approximations to the complementary error function have also been found,
including those where either only additions, multiplications, and one division,
or also the exponential, are used to reach either a uniform absolute, or a uniform relative, accuracy.}
The coefficients have been tabulated for up to 128-bit precision.
Tests of a computer code implementation
using the standard single- and double-precision floating-point arithmetic
show good performance and vectorizability.
\end{abstract}

\maketitle

\section{Introduction}

The error function~\cite{G871} of real argument
\begin{equation}
\label{eq:erf}
\erf(x) = \frac2{\sqrt{\pi}} \int_0^x \exp\left(-t^2 \right) \de t
\end{equation}
shows up in many mathematical models of physical and other phenomena
(far too many even to be listed here),
and its numerical evaluation can be a bottleneck of a computational simulation.
In some physical models~\cite{E21,DYP93,B50,GAP96} the error function divided by its argument ---
a well-behaved even function
\begin{equation}
\label{eq:erf0}
f_0(x) = \frac{\erf(x)}{x}
\end{equation}
should be carefully evaluated.
The complementary error function
\begin{equation}
\label{eq:erfc}
\erfc(x) = 1 - \erf(x)
\end{equation}
may need to be computed for $x \ge 0$ in some cases,
and not by a simple subtraction from 1 with a loss of precision
(while for $x < 0$ it is right to do so).

Standard mathematical libraries of C and FORTRAN implement the functions~(\ref{eq:erf}) and~(\ref{eq:erfc}) since at least 2008,
and use diverse approximations for some defined ranges of $x$, favoring accuracy over speed,
but it can be helpful to have a faster though slightly less accurate implementaion.
The vector instructions of modern processors promise a speedup of up to 16 times,
but then a branch-free code is needed to harness them.

We have found global closed-form approximations
to both functions~(\ref{eq:erf}) and~(\ref{eq:erf0}),
in terms of addition, multiplication, division, and square root,
with or without also using the exponential function
\mm{--- and likewise also to function~(\ref{eq:erfc}) ---}
where the accuracy can be systematically improved
by taking more polynomial terms with optimized coefficients,
reaching up to 128 bits of precision and stopping there.

We confess having found our approximation formulas
by general mathematical arguments using the natural intelligence of our own mind,
but then we had to make sure this had not been done before.
We see in the literature that the approximations to the error function
have been developed since the early days of computation~\cite{H55,H57,H66,O68,C69,SL81},
but ours still seem to be new.
(We cannot review all such works here as it may grow into a study in the psychology of mathematics\mm{;
``denn das \textit{Wesen} der \textit{Mathematik} liegt gerade in ihrer \textit{Freiheit}''\cite{C83}}.)

\section{Approximations}

We begin with a transformation of the error function~(\ref{eq:erf})
\begin{equation}
\label{eq:erfa}
\erf(x) = \frac{x}{\sqrt{x^2 + \phi\left(x^2 \right)}}
\end{equation}
in terms of a new function $\phi(s)$,
the $x$ in the numerator in~(\ref{eq:erfa}) makes it ideal also for the function~(\ref{eq:erf0}).
Looking at its explicit form
\begin{equation}
\phi(s) = \frac{s}{\bigl[\erf\bigl(\sqrt{s}\bigr) \bigr]^2} - s,
\end{equation}
one may be misled into thinking it is not good for approximations, but it is.
We need $\phi(s)$ only for $s\ge 0$ where it is monotonically decreasing,
starting from
\begin{equation}
\label{eq:phi0}
\phi(0) = \frac{\pi}4,
\end{equation}
with the negative first derivative
\begin{equation}
\label{eq:phi1}
\phi'(0) = \frac{\pi}{6} - 1,
\end{equation}
and all the way to the asymptotic limit
\begin{equation}
\label{eq:phi8}
\lim_{s\to\infty} \phi(s) = \frac2{\sqrt{\pi}} \sqrt{s} \exp(-s) .
\end{equation}
It is natural to further transform
\begin{equation}
\label{eq:re}
\phi(s) = \sqrt{\psi(s)}\,\exp(-s),
\end{equation}
so that for the new function $\psi(s)$ the rational approximation
\begin{equation}
\label{eq:qn}
\psi_N(s) = \frac{\sum_{m=0}^{N + 1} A_{mN} s^m}{1 + \sum_{n=1}^N B_{nN} s^n} \approx \psi(s)
\end{equation}
can be made.
The conditions~(\ref{eq:phi0}),~(\ref{eq:phi1}), and~(\ref{eq:phi8}) now become
\begin{equation}
\label{eq:psi0}
\psi(0) = \frac{\pi^2}{16},
\end{equation}
\begin{equation}
\label{eq:psi1}
\psi'(0) = \frac{(5\pi - 12) \pi}{24},
\end{equation}
\begin{equation}
\label{eq:psi8}
\lim_{s\to\infty} \psi(s) = \frac4{\pi} s,
\end{equation}
and the rational function~(\ref{eq:qn}) can be easily constrained to satisfy them.
One more asymptotic term
\begin{equation}
\label{eq:psi9}
\lim_{s\to\infty} \left(\psi(s) - \frac4{\pi} s \right) = -\frac4{\pi}
\end{equation}
could also be used when working with $\erfc$,
while it would be of little help for $\erf$.

Knowing that the exponential function of negative argument can be approximated,
to a given uniform absolute accuracy, by the expression
\begin{equation}
\exp(-s) \approx \textstyle \left( 1 + \sum_{n=1}^{N} (2^{-K} s)^n b_n /n! \right)^{-2^K}
\end{equation}
with either exact $b_n = 1$ or optimized $b_n \approx 1$, and with the right $N$ and $K$,
we have also sought the approximations to the error function, to a given relative accuracy,
in terms of arithmetic and square root operations only.
We have ended up finding the approximations
\begin{equation}
\label{eq:pmnk}
\phi_{MN}^{(K)}(s) =
\left(\frac{\sum_{m=0}^M A_{mMN}^{(K)} s^m}{1 + \sum_{n=1}^N B_{nMN}^{(K)} s^n}\right)^{2^K}
\approx \phi(s)
\end{equation}
to work strikingly well for the right $M$, $N$, and $K$,
even without satisfying~(\ref{eq:phi8}).
\mm{(Approximations to other sigmoid functions can be made using Eqs.~(\ref{eq:erfa}) and~(\ref{eq:pmnk}) ---
for $\tanh(x)$, $K = 1$, and $M = 0$ there is an analytic expression $B_n = 1/(2n + 1)!$ based on Taylor series.)}

Approximations to the complementary error function~(\ref{eq:erfc}) for $x \ge 0$ can also be computed,
instead of subtracting $1$ from~\ref{eq:erfa},
by a numerically stable expression
\begin{equation}
\erfc(x) = \frac{\phi(x^2)}{x^2 + \phi(x^2) + x\sqrt{x^2 + \phi(x^2)}}.
\end{equation}
Uniform relative accuracy for $\erfc(x)$ can be reached so far
only by the exponential-based approximation~(\ref{eq:re})
with the reoptimized coefficients in Eq.~(\ref{eq:qn})
and nearly twice as many $N$ polynomial terms are needed
for the same accuracy as for $\erf(x)$.

\mm{
To our own surprise, the ``simplest'' approximation
\begin{equation}
\label{eq:ern}
f_N(x) = \frac{1 + \sum_{m=1}^N C_{mN} x^m}{1 + \sum_{n=1}^{N + 1} D_{nN} x^n} \exp\left(-x^2\right) \approx \erfc(x),
\end{equation}
for all $x\ge 0$, works so well it is hard to believe it was overlooked.
The coefficients in Eq.~(\ref{eq:ern}) can be optimized to get approximations
with uniform absolute as well as relative accuracy,
for the latter the constraint
\begin{equation}
\label{eq:ercn}
D_{N + 1, N} = \sqrt{\pi} C_{NN}
\end{equation}
should be used to fix the $x\to\infty$ asymptotics,
and we always get well-behaved solutions with all positive $\{C_{mN}\}$ and $\{D_{nN}\}$ for all $N$ studied.
In other words, the exponentially scaled complementary error function
\begin{equation}
\label{eq:erfcx}
\erfcx(x) = \erfc(x) \exp\left(x^2\right)
\end{equation}
is a great example of a function for which the rational approximation works well.
}

For a uniform absolute accuracy, Eq.~(\ref{eq:pmnk}) could be used,
but we have suddenly found even simpler direct approximations
\begin{equation}
\label{eq:fmnk}
f_{MN}^{(P)}(x) =
\left(\frac{1 + \sum_{m=1}^M C_{mMN}^{(P)} x^m}{1 + \sum_{n=1}^N D_{nMN}^{(P)} x^n}\right)^{P}
\approx \erfc(x)
\end{equation}
working well with the right $M$, $N$, and $P = 2^K$
(a few special cases with $M = 0$ were known~\cite{H55} before).
In the applications~\cite{DYP93} where both $\erfc(x)$ and its derivative are needed,
it may be more consistent to compute the derivative of the approximation itself ---
it is then much better to take $P = 2^K + 1$ in Eq.~(\ref{eq:fmnk}) for speed's sake.

To get a uniform approximation $f(x)$ of an exact function $F(x)$,
the coefficients should be optimized to minimize the maximum
\begin{equation}
\label{eq:em}
E = \max\limits_{0 < x < \infty} \bigl|\varepsilon(x) \bigr|
\end{equation}
(weighted) error
\begin{equation}
\varepsilon(x) = \bigl(f(x) - F(x)\bigr) w(x).
\end{equation}
A uniform relative accuracy is reached with the weight
\begin{equation}
\label{eq:wf}
w(x) = \frac1{F(x)},
\end{equation}
and only a uniform absolute accuracy with $w(x) = 1$.
We also find a kind of interpolation between these ends with
\begin{equation}
\label{eq:wh}
w(x) = \frac1{\sqrt{h^2 + \bigl(F(x)\bigr)^2}}
\end{equation}
to be useful for functions like $F(x) = \erfc(x)$,
and a self-consistent choice of $h$ seems to be
\begin{equation}
\label{eq:eh}
h = E(h),
\end{equation}
\mm{or $h$ can be set to the smallest representable (denormal or not) number in the target floating-point format.}

In practice, the minimization of~(\ref{eq:em})
can be done by solving the system of equations
\begin{equation}
\label{eq:ex}
\left\{\begin{array}{ccrll}
\varepsilon(x_i) &=& -\varepsilon(x_{i+1}), & x_i < x_{i+1}, & i=1,\dots,L, \\
\varepsilon'(x_i) &=& 0, & & i=1,\dots,L + 1,
\end{array}\right.
\end{equation}
for $L$ variables (the approximation coefficients).
One more variable $h$ and one more equation
\begin{equation}
\bigl|\varepsilon(x_1)\bigr| = h
\end{equation}
can be used when working with~(\ref{eq:wh}) and~(\ref{eq:eh}).

The starting values can be taken first from the direct minimization
of the least-squares ($p=1$)
\begin{equation}
\label{eq:ep}
E^{(p)} = \int\limits_0^\infty \bigl(\varepsilon(x)\bigr)^{2p} \de x + h^{2p}
\end{equation}
or more general ($p=2,3,\dots$) functional.

We get the most desired uniform relative accuracy
for an approximation of $\erf(x)$ by $f(x)$ based on~(\ref{eq:erfa})
with $\phi(s)$ either from~(\ref{eq:re}) and~(\ref{eq:qn}) or from~(\ref{eq:pmnk}),
using~(\ref{eq:wf}) and solving Eqs.~(\ref{eq:ex})
for $L$ variables: $L=2N-1$ for~(\ref{eq:qn})
with~(\ref{eq:psi0}), (\ref{eq:psi1}), and~(\ref{eq:psi8});
or $L=M+N-1$ for~(\ref{eq:pmnk}) with~(\ref{eq:phi0}) and~(\ref{eq:phi1}).
\mm{For $\erfc(x)$, Eqs.~(\ref{eq:ern}) and~(\ref{eq:ercn}) make $L = 2N$, and $L = M + N$ for Eq.~(\ref{eq:fmnk}).}

\section{Computations}

We have written a computer code to determine the approximation coefficients
using multiple-precision floating-point arithmetic, typically 256 bits.
Table~\ref{tab:e} shows the accuracy of these approximations
given as $\epsilon = -\log_2 E$, the number of significant bits,
when the computation is done with a much higher bit precision.

\begingroup
\begin{table*}[h]
\caption{\label{tab:e}Accuracy of approximations.}
\begin{ruledtabular}
\begin{tabular}{rrrr|rrr|rrrr|rrrr}
\multicolumn{7}{c|}{\it exponential-based} & \multicolumn{8}{c}{\it exponential-free} \\
\multicolumn{4}{c}{Eq.~(\ref{eq:qn})} & \multicolumn{3}{c|}{\mm{Eq.~(\ref{eq:ern})}} & \multicolumn{4}{c}{Eq.~(\ref{eq:pmnk})} & \multicolumn{4}{c}{Eq.~(\ref{eq:fmnk})} \\
\hline
 $N$ & \multicolumn{3}{c|}{$\epsilon$\footnotemark[1]} &\mm{$N$} & \mm{$\bar{\epsilon}$} & \mm{$\epsilon$}
                                 & $M$   & $N$   & $K$  & $\epsilon$\footnotemark[1]
                                                                  & $M$   & $N$   & $K$  & $\bar{\epsilon}$\footnotemark[2] \\
\hline
     & $\erf$ &\multicolumn{2}{c|}{$\erfc$\footnotemark[3]} & &\multicolumn{2}{c|}{\mm{$\erfc$}} & & & & $\erf$ & & & & $\erfc$ \\
\cline{3-4}\cline{6-7}
     1 &    11.0 &  5.9 &         &\mm{     1}&\mm{     15.0}&\mm{      8.2}&     0 &     3 &    1 &    11.5 &    0 &     4 &     2 &    11.1 \\
     2 &    17.6 &  9.8 &     8.8 &\mm{\bf  2}&\mm{\bf  22.9}&\mm{     13.6}&     0 &     4 &    2 &    16.7 &    0 &     4 &     5 &    15.4 \\
\bf  3 &\bf 24.2 & 13.4 &    12.7 &\mm{     3}&\mm{     30.8}&\mm{     19.0}&\bf  0 &\bf  5 &\bf 2 &\bf 22.7 &\bf 0 &\bf  7 &\bf  4 &\bf 23.5 \\
     4 &    29.9 & 16.8 &    16.2 &\mm{\bf  4}&\mm{     38.6}&\mm{\bf  24.3}&     3 &     5 &    6 &    29.6 &    2 &     6 &     7 &    29.0 \\
     5 &    34.0 & 20.3 &    19.8 &\mm{     5}&\mm{     46.5}&\mm{     29.7}&     2 &     8 &    3 &    33.8 &    2 &     8 &     6 &    34.5 \\
\bf  6 &    40.5 & 23.6 &\bf 23.2 &\mm{\bf  6}&\mm{\bf  54.3}&\mm{     35.0}&     3 &    10 &    3 &    40.2 &    3 &     8 &     8 &    39.4 \\
     7 &    42.4 & 26.8 &    26.5 &\mm{     7}&\mm{     62.2}&\mm{     40.4}&     5 &     8 &    5 &    41.9 &    4 &     9 &     8 &    45.8 \\
     8 &    48.3 & 30.5 &    30.2 &\mm{     8}&\mm{     70.0}&\mm{     45.7}&     4 &    12 &    3 &    47.4 &    4 &     9 &     9 &    49.4 \\
\bf  9 &\bf 53.9 & 33.3 &    33.1 &\mm{\bf  9}&\mm{     77.9}&\mm{\bf  51.1}&\bf  6 &\bf 10 &\bf 5 &\bf 52.2 &\bf 5 &\bf 10 &\bf 10 &\bf 53.3 \\
    10 &    60.1 & 35.6 &    35.5 &\mm{    10}&\mm{     85.7}&\mm{     56.5}&     8 &    12 &    6 &    57.9 &    6 &    11 &     9 &    59.0 \\
    11 &    62.0 &      &    39.1 &\mm{    11}&\mm{     93.6}&\mm{     61.8}&     8 &    13 &    5 &    58.6 &    6 &    11 &    11 &    61.0 \\
    12 &    64.7 & 40.6 &    40.5 &\mm{    12}&\mm{    101.4}&\mm{     67.2}&     9 &    14 &    6 &    64.0 &    6 &    12 &    10 &    63.1 \\
    13 &    70.7 &      &\ts{43.7}&\mm{    13}&\mm{    109.3}&\mm{     72.6}&    10 &    15 &    6 &    68.3 &    7 &    13 &     9 &    68.0 \\
    14 &    76.0 &      &\ts{46.3}&\mm{    14}&\mm{    117.1}&\mm{     78.0}&    11 &    17 &    5 &    75.0 &    8 &    15 &    10 &    78.8 \\
    15 &    81.0 &      &\ts{49.3}&\mm{    15}&\mm{    125.0}&\mm{     83.3}&    13 &    18 &    6 &    80.4 &    8 &    16 &    10 &    82.1 \\
    16 &    86.3 &      &\ts{52.6}&\mm{    16}&\mm{    132.8}&\mm{     88.7}&    14 &    20 &    6 &    88.6 &    9 &    17 &    10 &    88.2 \\
    17 &         &      &\ts{56.0}&\mm{    17}&\mm{    140.7}&\mm{     94.1}&       &       &      &         &      &       &       &         \\
\bf 18 &    91.0 & 58.2 &\bf 58.2 &\mm{    18}&\mm{         }&\mm{     99.5}&    15 &    20 &    6 &    90.6 &    9 &    17 &    11 &    89.6 \\
    19 &    96.5 &      &\ts{62.7}&\mm{    19}&\mm{         }&\mm{    104.9}&    17 &    20 &    8 &    93.5 &   10 &    17 &    12 &    92.7 \\
    20 &   101.2 & 63.3 &\ts{63.3}&\mm{    20}&\mm{         }&\mm{    110.3}&    17 &    22 &    8 &    99.8 &   11 &    18 &    12 &    98.3 \\
    21 &         &      &\ts{64.1}&\mm{    21}&\mm{         }&\mm{    115.6}&    17 &    23 &    7 &   102.3 &   11 &    19 &    12 &   100.6 \\
    22 &   108.1 &      &\ts{68.3}&\mm{    22}&\mm{         }&\mm{    121.0}&    19 &    25 &    7 &   106.3 &   11 &    21 &    11 &   106.4 \\
    23 &   113.6 &      &\ts{70.9}&\mm{    23}&\mm{         }&\mm{    126.4}&    22 &    28 &    8 &   117.2 &   13 &    21 &    12 &   113.6 \\
    24 &   119.8 &      &\ts{74.2}&\mm{    24}&\mm{         }&\mm{    131.8}&    23 &    28 &    8 &   121.5 &   13 &    23 &    11 &   118.7 \\
    25 &         &      &\ts{77.4}&\mm{    25}&\mm{         }&\mm{         }&       &       &      &         &      &       &       &         \\
    26 &   125.4 &      &\ts{80.1}&\mm{    26}&\mm{         }&\mm{         }&    25 &    30 &    8 &   123.2 &   14 &    24 &    11 &   122.1 \\
    27 &   130.1 &      &\ts{84.6}&\mm{    27}&\mm{         }&\mm{         }&    25 &    31 &    8 &   130.2 &   16 &    24 &    12 &   129.5 \\
\end{tabular}
\end{ruledtabular}
\footnotetext[1]{The maximum \textit{relative} error of approximation $E$ of Eq.~(\ref{eq:em})
is given as the negative binary logarithms $\epsilon = -\log_2 E$.}
\footnotetext[2]{The maximum \textit{absolute} error of approximation $\bar{E}$ of Eq.~(\ref{eq:em})
is given as the negative binary logarithms $\bar{\epsilon} = -\log_2 \bar{E}$.}
\footnotetext[3]{The two columns for $\erfc$ are without and with the condition of Eq.~(\ref{eq:psi9}),
values in brackets are for solutions having some negative coefficients.}
\end{table*}
\endgroup

For the exponential-based approximation~(\ref{eq:qn})
we have found well-behaved solutions, with all coefficients positive, for up to $N=27$,
but failed for $N=17,21,25$ where some $B_{nN}<0$,
and we see an exponential convergence with $N$.

For the exponential-free approximation~(\ref{eq:pmnk}) we do not claim
to have worked through all combinations of $(M,N,K)$,
nevertheless we have found 55 well-behaved solutions
some of which are shown in Table~\ref{tab:e}
alongside the exponential-based solutions of comparable accuracy.

Remarkably, both approximations need almost the same number of polynomial terms
to reach a given accuracy.
Thus the latter can be faster
as $K$ multiplications are faster than the exponential function,
but the former is still useful if the values of both $\erf(x)$ and $\exp\left(-x^2 \right)$ are needed.

The complementary error function is harder (Table~\ref{tab:e}),
the exponential-based approximation of Eq.~(\ref{eq:qn})
needs about twice as many terms for the same relative accuracy
and many solutions have some negative coefficients,
but luckily that of Eq.~(\ref{eq:ern}) works so well it can be called ``ideal''.
The simplest exponential-free approximation~(\ref{eq:fmnk})
with uniform absolute accuracy also behaves well.

To study the effects of finite-precision arithmetic,
and also as a way to share all our solutions,
we have formatted the coefficients as C code~(see \href{https://}{supplementary material})
to evaluate the approximations
in 24-bit (mantissa) single, 53-bit double, 64-bit long double, and 113-bit quadruple precision,
and to compare it to the standard library $\erf$ functions,
and also a separate code using the MPFR~\cite{FHLPZ07,mpfr} library.
As our ``standard'' single- and double-precision approximations
we have chosen those highlighted in Table~\ref{tab:e},
and the rounding errors add up to leave us
with about 22, 21 (single) and 51, 48 (double) bits of precision for $\erf$.

To measure the computational speed, we have written C code (see \href{https://}{supplementary material})
for serial, 4- and 8-way double-, and 8- and 16-way single-precision vectorized calculation,
such that the GCC~\cite{gcc} compiler we use can translate it into either scalar or vector instructions.
For the AVX2/FMA and AVX512F instruction sets, we get a quite well-optimized machine code
where the instructions in scalar and vector versions nearly parallel each other.
We run it on two 16-core processors: an AMD 3950X (from 2019) clocked at 3.5 GHz
and an AMD 9950X (from 2024) clocked at 4.3 GHz,
both with SMT (simultaneous multithreading) disabled,
16 identical jobs in parallel to load all the cores.
Timing the repeated evaluation of a function $f(x)$ over 512 equally-spaced values of $0\le x <4$,
for a total of about $2^{32}$ function calls,
is used to estimate the number of processor clock cycles
for one function value including load/store, call/return, and looping operations.

We compare the speed of the standard C library~\cite{glibc} implementation of $\exp$ and $\erf$ functions
against the serial and vectorized code of our approximations.
We also use this occasion to share our own vectorized single- and double-precision implementation
of the exponential function where not the traditional Chebyshev
but the direct uniform approximation
to the $2^x$ function for $-\tfrac12 \le x \le \tfrac12$ is used.

\begingroup
\begin{table*}[h]
\caption{\label{tab:t}Measurements of computational speed.}
\begin{ruledtabular}
\begin{tabular}{lll|rrrrrrr|rrrrrrr}
function            & method                & $\varepsilon$\footnotemark[1]
                                                   &        & \multicolumn{2}{c}{clock}  & \multicolumn{4}{c|}{speedup}
                                                   &        & \multicolumn{2}{c}{clock}  & \multicolumn{4}{c}{speedup} \\
\cline{7-10}
\cline{14-17}
                    &                       &
                                     &$m$\footnotemark[2]  & \multicolumn{2}{c}{cycles} & \multicolumn{2}{c}{library} & \multicolumn{2}{c|}{vectorize}
                                     &$m$\footnotemark[2]  & \multicolumn{2}{c}{cycles} & \multicolumn{2}{c}{library} & \multicolumn{2}{c}{vectorize} \\
\hline
&&&& Z2\footnotemark[3] & Z5\footnotemark[4] & Z2  & Z5  & Z2  & Z5
                                                                                                                    & & Z2 & Z5 & Z2 & Z5 & Z2 & Z5   \\
&&&\multicolumn{3}{r}{\it double-precision} &&&&
                                                                                                                    &\multicolumn{3}{r}{\it single-precision} &&&& \\
    $\exp$          &    glibc~\cite{glibc} &\mm{r}&     1 &     17.0 &     12.1 &        &        &        &         &     1 &      9.1 &      8.1 &        &        &        &         \\
    $\exp$          &    ours               &\mm{r}&     1 &     15.7 &      9.2 &    1.1 &    1.3 &        &         &     1 &      9.9 &      6.0 &    0.9 &    1.4 &        &         \\
    $\exp$          &    ours               &\mm{r}&     4 &     18.5 &     10.3 &        &        &    3.4 &     3.6 &     8 &     11.4 &      6.9 &        &        &    6.9 &     7.0 \\
    $\exp$          &    ours               &\mm{r}&     8 &          &     10.5 &        &        &        &     7.0 &    16 &          &      6.8 &        &        &        &    14.1 \\
    $\erf$          &    glibc~\cite{glibc} &\mm{r}&     1 &     50.7 &     30.5 &        &        &        &         &     1 &     23.2 &     15.1 &        &        &        &         \\
    $\erf$, $\exp$  &    Eq.~\ref{eq:qn}    &\mm{r}&     1 &     42.8 &     27.2 &        &        &        &         &     1 &     26.7 &     17.8 &        &        &        &         \\
    $\erf$, $\exp$  &    Eq.~\ref{eq:qn}    &\mm{r}&     4 &     65.2 &     28.1 &        &        &    2.6 &     3.9 &     8 &     37.6 &     23.6 &        &        &    5.7 &     6.0 \\
    $\erf$, $\exp$  &    Eq.~\ref{eq:qn}    &\mm{r}&     8 &          &     28.4 &        &        &        &     7.7 &    16 &          &     30.1 &        &        &        &     9.5 \\
    $\erf$          &    Eq.~\ref{eq:pmnk}  &\mm{r}&     1 &     26.7 &     17.9 &    1.9 &    1.7 &        &         &     1 &     14.7 &      9.2 &    1.6 &    1.6 &        &         \\
    $\erf$          &    Eq.~\ref{eq:pmnk}  &\mm{r}&     4 &     34.0 &     18.0 &        &        &    3.1 &     4.0 &     8 &     20.4 &     12.1 &        &        &    5.8 &     6.1 \\
    $\erf$          &    Eq.~\ref{eq:pmnk}  &\mm{r}&     8 &          &     18.0 &        &        &        &     8.0 &    16 &          &     18.2 &        &        &        &     8.1 \\
    $\erfc$         &    glibc~\cite{glibc} &\mm{r}&     1 &     53.2 &     31.8 &        &        &        &         &     1 &     49.7 &     29.3 &        &        &        &         \\
    $\erfc$, $\exp$ &    Eq.~\ref{eq:qn}    &\mm{r}&     1 &     74.1 &     45.5 &        &        &        &         &     1 &     31.2 &     20.6 &        &        &        &         \\
    $\erfc$, $\exp$ &    Eq.~\ref{eq:qn}    &\mm{r}&     4 &    102.4 &     47.1 &        &        &    2.9 &     3.9 &     8 &     60.8 &     36.0 &        &        &    4.1 &     4.6 \\
    $\erfc$, $\exp$ &    Eq.~\ref{eq:qn}    &\mm{r}&     8 &          &     45.9 &        &        &        &     7.9 &    16 &          &     38.6 &        &        &        &     8.5 \\
\mm{$\erfc$, $\exp$}&\mm{Eq.~\ref{eq:ern}  }&\mm{r}&\mm{ 1}&\mm{ 34.8}&\mm{ 21.5}&\mm{1.5}&\mm{1.5}&\mm{   }&\mm{    }&\mm{ 1}&\mm{ 19.9}&\mm{ 12.6}&\mm{2.5}&\mm{2.3}&\mm{   }&\mm{    }\\
\mm{$\erfc$, $\exp$}&\mm{Eq.~\ref{eq:ern}  }&\mm{r}&\mm{ 4}&\mm{ 40.0}&\mm{ 22.6}&\mm{   }&\mm{   }&\mm{3.5}&\mm{ 3.8}&\mm{ 8}&\mm{ 25.2}&\mm{ 15.0}&\mm{   }&\mm{   }&\mm{6.3}&\mm{ 6.7}\\
\mm{$\erfc$, $\exp$}&\mm{Eq.~\ref{eq:ern}  }&\mm{r}&\mm{ 8}&\mm{     }&\mm{ 22.8}&\mm{   }&\mm{   }&\mm{   }&\mm{ 7.5}&\mm{16}&\mm{     }&\mm{ 15.6}&\mm{   }&\mm{   }&\mm{   }&\mm{12.9}\\
\mm{$\erfcx$       }&\mm{Eq.~\ref{eq:ern}  }&\mm{r}&\mm{ 1}&\mm{ 15.0}&\mm{ 10.6}&\mm{   }&\mm{   }&\mm{   }&\mm{    }&\mm{ 1}&\mm{  8.1}&\mm{  5.5}&\mm{   }&\mm{   }&\mm{   }&\mm{    }\\
\mm{$\erfcx$       }&\mm{Eq.~\ref{eq:ern}  }&\mm{r}&\mm{ 4}&\mm{ 20.2}&\mm{ 10.8}&\mm{   }&\mm{   }&\mm{3.0}&\mm{ 3.9}&\mm{ 8}&\mm{ 10.8}&\mm{  6.5}&\mm{   }&\mm{   }&\mm{6.0}&\mm{ 6.8}\\
\mm{$\erfcx$       }&\mm{Eq.~\ref{eq:ern}  }&\mm{r}&\mm{ 8}&\mm{     }&\mm{ 11.1}&\mm{   }&\mm{   }&\mm{   }&\mm{ 7.6}&\mm{16}&\mm{     }&\mm{  6.3}&\mm{   }&\mm{   }&\mm{   }&\mm{14.0}\\
    $\erfc$         &    Eq.~\ref{eq:fmnk}  &\mm{a}&     1 &     27.1 &     16.3 &    2.0 &    2.0 &        &         &     1 &     11.1 &      7.4 &    4.5 &    4.0 &        &         \\
    $\erfc$         &    Eq.~\ref{eq:fmnk}  &\mm{a}&     4 &     27.9 &     16.5 &        &        &    3.9 &     4.0 &     8 &     16.1 &      9.1 &        &        &    5.5 &     6.5 \\
    $\erfc$         &    Eq.~\ref{eq:fmnk}  &\mm{a}&     8 &          &     16.5 &        &        &        &     7.9 &    16 &          &      9.1 &        &        &        &    13.0 \\
    $\erfc$, $\exp$ &    Eq.~\ref{eq:fmnk}  &\mm{a}&     1 &     45.6 &     28.7 &        &        &        &         &     1 &     19.9 &     13.1 &        &        &        &         \\
    $\erfc$, $\exp$ &    Eq.~\ref{eq:fmnk}  &\mm{a}&     4 &     46.4 &     28.1 &        &        &    3.9 &     4.1 &     8 &     24.4 &     14.2 &        &        &    6.5 &     7.4 \\
    $\erfc$, $\exp$ &    Eq.~\ref{eq:fmnk}  &\mm{a}&     8 &          &     28.4 &        &        &        &     8.1 &    16 &          &     14.5 &        &        &        &    14.5 \\
\end{tabular}
\end{ruledtabular}
\footnotetext[1]{\mm{Accuracy, relative (r) or absolute (a).}}
\footnotetext[3]{\mm{Vector length $m$.}}
\footnotetext[3]{Z2: AMD 3950X}
\footnotetext[4]{Z5: AMD 9950X}
\end{table*}
\endgroup

Table~\ref{tab:t} shows our measurements,
we see a not-so-unexpected speedup against the standard library,
and a rather good vectorization speedup ---
less than ideal because, among other things,
the processor shows greater superscalar capabilities when fed with a scalar instruction stream.
The newer 9950X processor shows an ideal 4- and 8-way double-precision speedup
for our $\erf$ approximation, while a somewhat poor single-precision performance
has a clear cause: the lack of a full support for the reciprocal and square root operations
in the single-precision vector instruction set,
instead only 14-bit approximations are computed
and then refined by one Newton-Raphson iteration.

\section{Conclusions}

We have found two new kinds of global closed-form approximations to the error function of real argument,
and \mm{two} more for the complementary error function of non-negative real argument,
and determined their coefficients and accuracy.
The number of terms needed to reach an accuracy of up to 128 bits is rather small.
Tests of a practical implementation using the (24-bit) single- and (53-bit) double-precision arithmetic
show a speed high enough to outperform on average a standard library routine in serial computation,
whereas the code is straightforward to vectorize and shows then a close-to-ideal performance.

The exponential-based functional form can be generalized~\cite{L25}
for the approximation of Boys~\cite{B50} functions
related to the lower incomplete gamma function of half-integer parameter.

\section{Supplementary Material}

See \href{https://doi.org/insert.later/}{supplementary material} for all approximation coefficients
and computer codes to test them.

\begin{acknowledgments}
The Author has received financial support from the Russian Science Foundation (Grant No. 26-16-00222).
\end{acknowledgments}

\section{Data availability}

The data that suppot the findings of this study are available within the article and its \href{https://}{supplementary material}.

The approximation coefficients and computer codes are also available online~\cite{erf26}.

\bibliography{erf}

\end{document}